\documentclass[12pt]{article}
\usepackage[top=1.5in, bottom=1.5in, left=1.25in, right=1.25in]{geometry}
\usepackage[english]{babel}
\usepackage{atbegshi,cite}
\usepackage{amsmath,amssymb,amsbsy,amstext, amsthm, simplewick}
\usepackage{hyperref}
\usepackage{graphicx}
\usepackage{amsfonts}
\usepackage[small]{caption}
\usepackage{upgreek}
\usepackage[titletoc]{appendix}
\usepackage{setspace}
\usepackage{color}
\usepackage{sectsty}

\sectionfont{\fontsize{14}{15}\selectfont}

\newcommand{\newc}{\newcommand}
\newc{\fpi}{f_{\pi}}
\newc{\etap}{\eta^{\prime}}
\newc{\llll}{\langle\lambda\lambda\rangle}
\newc{\FFd}{F^a\tilde F^a}
\newc{\qbar}{{\overline q}}
\newc{\TR}{{\rm Tr}}
\newc{\Kahler}{K\"ahler }
\newc{\Zbb}{{\mathbb Z}}
\newc{\Rd}{{\mathbb R}^d}
\newc{\Rt}{{\mathbb R}^3}
\newc{\Ro}{{\mathbb R}^1}
\newc{\Rf}{{\mathbb R}^4}
\newc{\So}{{\mathbb S}^1}
\newc{\zt}{{\mathbb Z}_2}
\newc{\RtSo}{{\mathbb R}^3\times{\mathbb S}^1}
\newc{\scriminus}{{\cal I}^-}
\newc{\scriplus}{{\cal I}^+}
\newc{\mpl}{M_p}
\newc{\Ricci}{\mathcal{R}}
\newc{\bv}{\phi}
\newc{\calU}{{\cal U}}
\newc{\calK}{K}
\newc{\calUi}{{\cal U}^{-1}}
\newc{\calG}{{\cal G}}
\newc{\calO}{{\cal O}}
\newc{\calOb}{{\cal O}^\dagger}
\newc{\calD}{{\cal D}}
\newc{\calA}{{\cal A}}
\newc{\calS}{{\cal S}}
\newc{\calL}{{\cal L}}
\newc{\hphi}{{\hat\phi}}

\begin{document}
\begin{titlepage}
\begin{flushright}
{\large 
~\\
}
\end{flushright}

\vskip 2.2cm

\begin{center}

{\large \bf Virtual and Thermal Schwinger Processes}

\vskip 1.4cm

{Patrick Draper}
\\
\vskip 1cm
{Department of Physics, University of Illinois, Urbana, IL 61801}\\
{Amherst Center for Fundamental Interactions, Department of Physics,\\ University of Massachusetts, Amherst, MA 01003}\\
\vspace{0.3cm}
\vskip 4pt

\vskip 1.5cm

\begin{abstract}
Electric flux may be screened by pair nucleation of heavy charges, a process that has a simple description in terms of a worldline instanton. When flux is wrapped around a small compact spatial dimension, worldline instantons  still induce  flux dissipation, but the leading process does not create real charged pairs. Instead, dissipation can be described in effective field theory as the production of long-wavelength scalar quanta via parametric resonance. The rate is computed semiclassically, and comments are made on the related problem of pair creation at finite temperature, for which differing results appear in the literature. Flux dissipation and the weak gravity conjecture together imply that the proper distance in field space a homogeneous axion field can traverse is bounded.
\end{abstract}

\end{center}

\vskip 1.0 cm

\end{titlepage}
\setcounter{footnote}{0} 
\setcounter{page}{1}
\setcounter{section}{0} \setcounter{subsection}{0}
\setcounter{subsubsection}{0}
\setcounter{figure}{0}

\section{Introduction}

Electric fields can be screened by the spontaneous creation of electron-positron pairs~\cite{Schwinger:1951nm}. In the weak, constant-field limit, $eE\ll m^2$, the pair production rate can be computed semiclassically in the worldline formalism. An instanton, in this case a Euclidean circle of radius $r_0=m/eE$, contributes an imaginary part to the effective action for the electromagnetic field of order $e^{-\pi m r_0}\ll 1$. Cutting the circle at the moment of time-symmetry and analytically continuing back to Minkowski space, the classical solution describes the nucleation of the charged pair at a separation 2$r_0$ and their subsequent acceleration away from each other in the background field.

In this note I revisit Schwinger pair production in the presence of a compact dimension, either a spatial or thermal circle. New semiclassical phenomena of lower action arise when the circumference of the circular dimension is less than $2r_0$. The primary focus here will be on the spatial case, which is the simpler of the two; subsequently I will comment on the thermal case, which has been a source of disagreement in the literature. For spatial circles, the lowest-action worldline instanton has winding number $\pm 1$ and has positive fluctuation determinant. At low energies, it generates a term in the effective action that carries an Aharonov-Bohm phase. Consequently, the instanton is still responsible for flux dissipation, proceeding through long-wavelength parametric resonance rather than production of heavy charged pairs.

The spatial case was previously discussed in~\cite{brown}. In terms of the instantons described here, the worldline studied in~\cite{brown} is an instanton-antiistanton event; it carries no phase and does not contribute to parametric resonance. Pair production at finite temperature was studied in~\cite{MO,brown}, which reached different conclusions, and more recently in~\cite{GR1,GR2,arun}. I will argue that the trajectories considered by~\cite{MO} in the thermal case are relevant to the spatial circle, while the trajectories in~\cite{brown} have some of the essential features needed in order to be associated with thermal pair production. 

In the spatial case, from the low energy point of view, the parametric resonance process describes the fragmentation of an axion field moving coherently with some velocity. Imposing the weak gravity conjecture on the mass of the charged matter, a bound is obtained on the maximum distance the axion can travel in field space.

\section{Worldline Instanton Amplitude}

We consider electrodynamics on $\Rf\times\So$ with circle circumference $L$ and a charged scalar of mass $m\gg1/L$, and focus on the terms in the Euler-Heisenberg action sensitive to $L$. 
Weak coupling is assumed for simplicity, $e\ll1$, and only constant electric flux in the compact dimension is turned on. I will also comment on the 1+1 dimensional case below. 

Integrating out the matter field, the Euclidean effective action  is 
\begin{equation}
\calS_{eff}=\calS_0-Z_1 
\end{equation}
where $\calS_0$ is the classical action, and the trace log correction can be expressed as a partition function of closed loops\footnote{For a textbook treatment, see~\cite{kleinert}.},
\begin{align}
Z_1&=-\TR\log(-D^2+m^2)\nonumber\\
&=\int_0^\infty \frac{d\beta}{\beta} \;e^{-\beta m/2}\int \calD x \;e^{-\calA}\,,
\end{align}
\begin{align}
\calA&=\int_0^\beta d\lambda \left(\frac{1}{2}m (\partial_\lambda x^M \partial_\lambda x_M) -i e A_M \partial_\lambda x^M\right)\nonumber\\
&=\int_0^\beta d\lambda \left(\frac{1}{2}m (\partial_\lambda x^M \partial_\lambda x_M) + e E_5 x^5 \partial_\lambda \tau \right)-ieA_5x^5(\lambda)\bigg|_0^\beta
\label{eq:Ainit}
\end{align}
Here $\tau=it$ is the Euclidean time and $x_5\sim x_5+L$ is the circle coordinate. In the last line, $A^0=0$ gauge is taken, and $A_5=A_5(\tau)$ with field strength $E_5=i\partial_\tau A_5$.

\begin{figure}[t!]
\begin{center}
\includegraphics[width=0.6\linewidth]{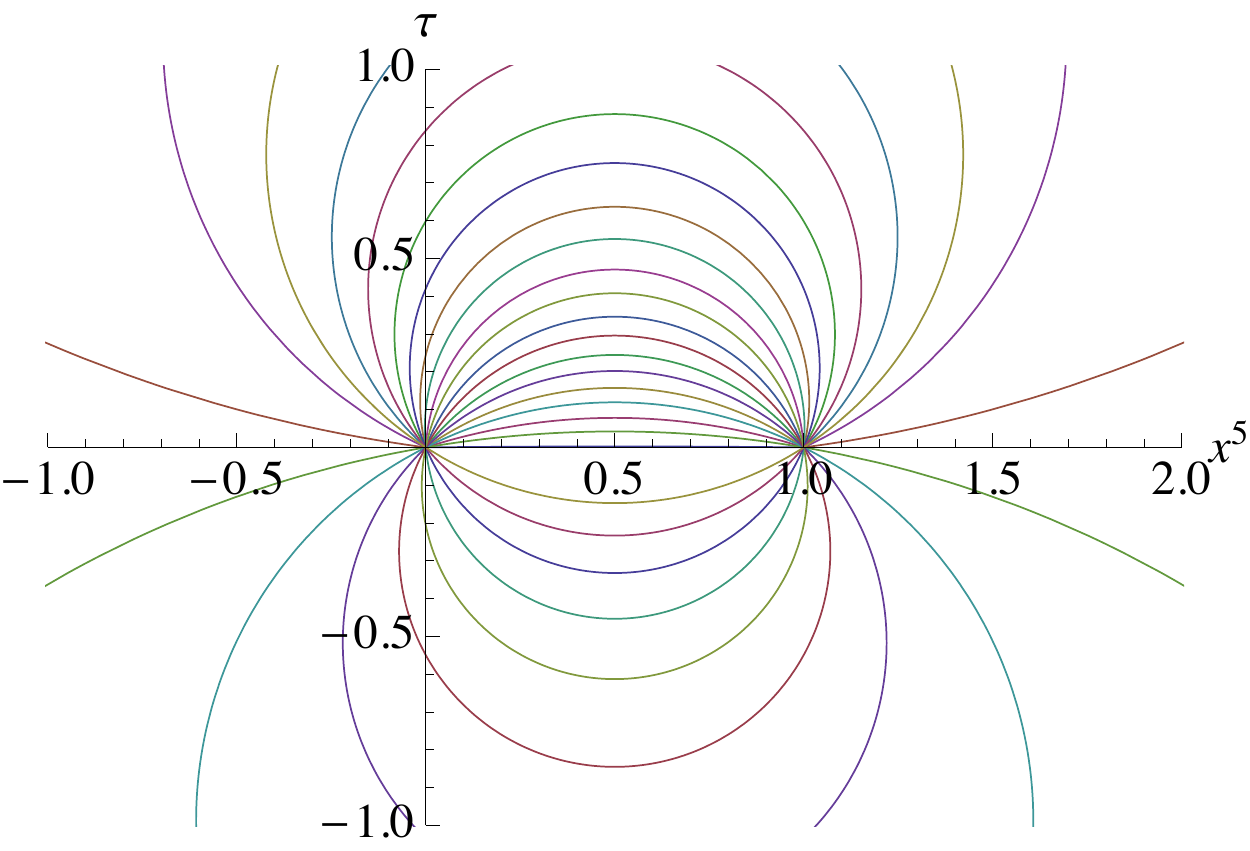}
\caption{Example Euclidean solutions to Eq.~(\ref{eq:xEOM}) for different values of $\beta$ ($L=1$, $r_0=1$, $x^5(\beta)=x^5(0)+L$)} 
\label{fig:sampletraj}
\end{center}
\end{figure}

The path integral over $x$ can be computed exactly for constant $E_5$ by expanding around classical solutions satisfying periodic boundary conditions. The leading $L$-dependent terms are obtained from solutions for which $x^5(\beta)-x^5(0)=\pm L$. The equations of motion are
\begin{align}
-m \partial_\lambda^2x_5+e E_5\partial_\lambda \tau&=0\nonumber\\
-m \partial_\lambda^2\tau-e E_5\partial_\lambda x_5&=0
\label{eq:xEOM}
\end{align}
The solutions are arcs of circles of angular size  $\int d\theta=\pm \beta/2r_0$. Examples are shown in Fig.~\ref{fig:sampletraj}. For angular lengths of magnitude greater than $2\pi$, the orbits appear as closed circles but end at $x^5(\beta)=x^5(0)\pm L$, indicating winding number $\pm 1$ around the fifth dimension.

The instantons are singular, but the classical action is finite,
\begin{equation}
\calA_{0}=\frac{1}{4}e E_5 L^2\cot\left(\omega\beta\right),\;\;\;\;\;\omega=1/2r_0=e E_5/2m\;.
\end{equation}
Since the action is quadratic, the fluctuations are insensitive to the classical solution,
\begin{equation}
\calA(\delta  x)=\int_0^\beta\left[\frac{1}{2}m(\partial_\lambda\delta x_M \partial_\lambda\delta x^M)+e E_5 \delta x^5 \partial_\lambda \delta\tau \right].
\end{equation}
The determinant is then the same as for solutions with zero winding number,
\begin{equation}
\left( \det \calO_{free}\right)^{-3/2}\left(\det \calO_{\tau x^5}\right)^{-1/2}
\end{equation}
with three powers of the free 1D particle partition function
\begin{equation}
\left(\det \calO_{free}\right)^{-1/2}\rightarrow V_1\sqrt{\frac{m}{2\pi\beta}}
\end{equation}
 and 
\begin{align}
\det \calO_{\tau x^5}&=\left(\det \calO_{free}\right)^2
\prod_{n\neq0} \det \left[\begin{matrix}
1 & \left(\frac{i\omega\beta}{\pi n}\right)\\
-\left(\frac{i\omega\beta}{\pi n}\right) & 1
\end{matrix}\right]\nonumber\\
&=\left(\det \calO_{free}\right)^2\prod_{n\neq0}\left[1-\left(\frac{\omega\beta}{\pi n}\right)^2\right]\nonumber\\
&=\left(\det \calO_{free}\right)^2\left(\frac{\sin(\omega\beta)}{\omega\beta}\right)^2\;.
\end{align}
The leading $L$-dependent contribution to $Z_1$ is therefore
\begin{equation}
Z^L_1=-2V_5\int_0^\infty \frac{d\beta}{\beta} \left(\frac{\omega\beta}{\sin(\omega\beta)}\right)\left(\frac{m}{2\pi\beta}\right)^{5/2}\;e^{-\frac{1}{2}m\left(\beta +\omega L^2\cot\left[\omega\beta\right]\right)}\cos(A_5 Le)
\label{eq:ZL1}
\end{equation}
with $\cos(A_5)$ dependence arising from the boundary term in Eq.~(\ref{eq:Ainit}) and the sum over $\Delta x^5=\pm L$.

\begin{figure}[t!]
\begin{center}
\includegraphics[width=0.45\linewidth]{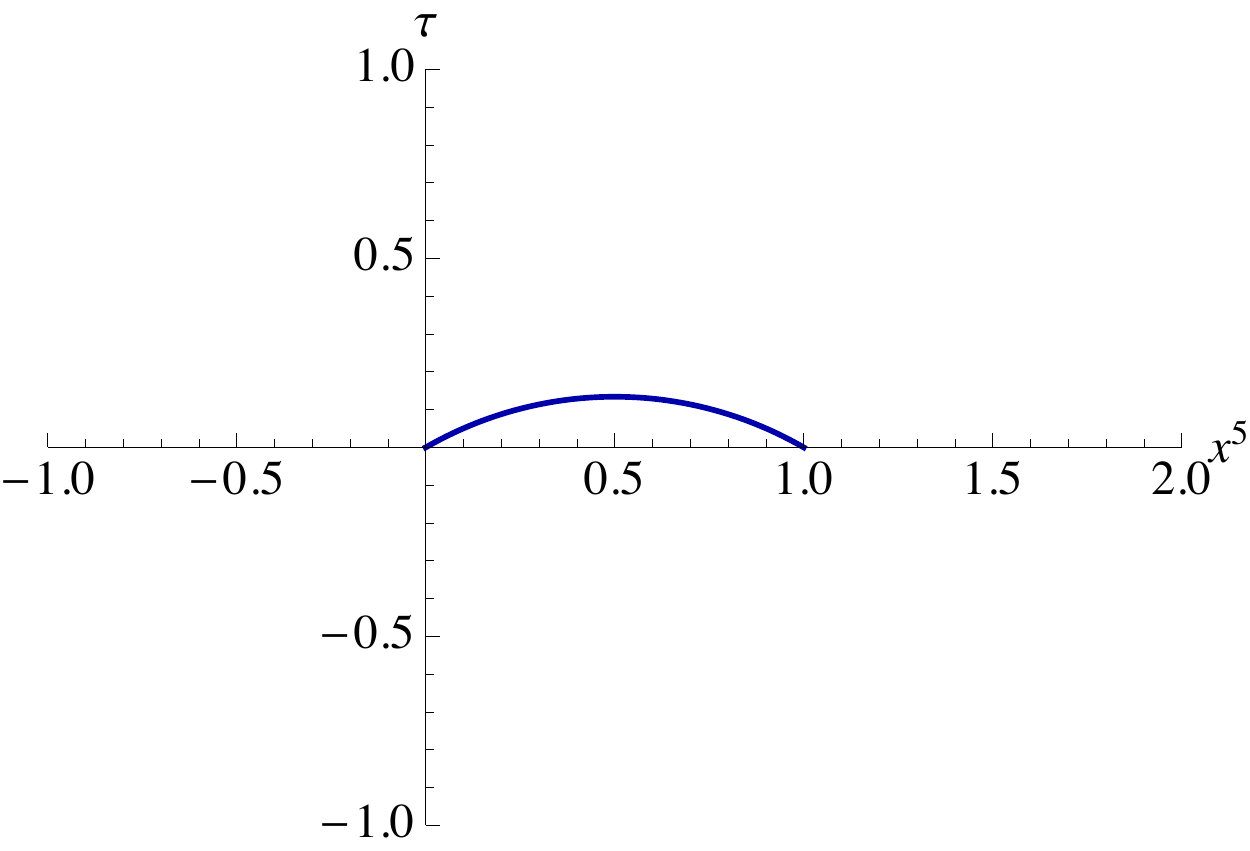}
\includegraphics[width=0.45\linewidth]{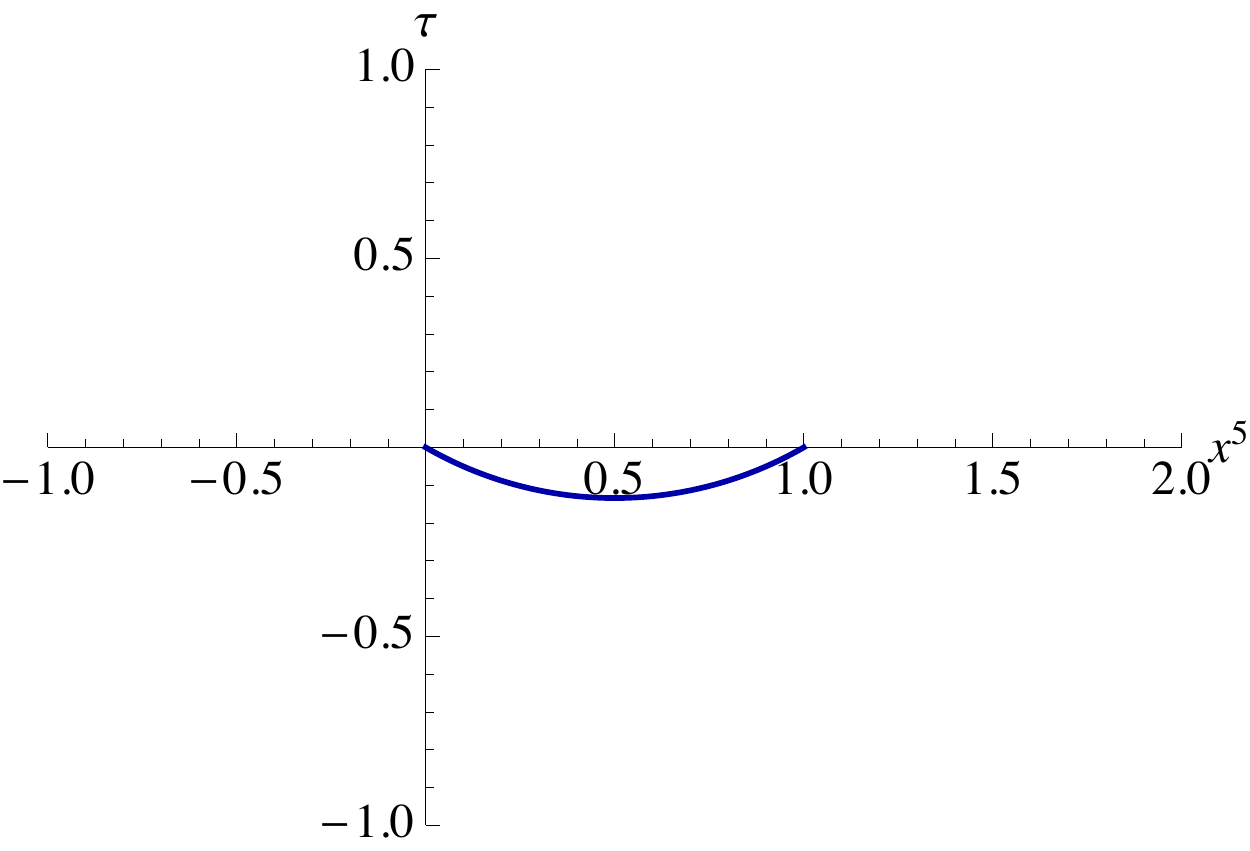}
\caption{Instanton with $\beta=\beta_0$. Parameters the same as in Fig.~\ref{fig:sampletraj}. Right: antiinstanton.}
\label{fig:beta0}
\end{center}
\end{figure} 

The contour of integration for $\beta$ must be chosen to avoid the singularities at $\beta=\pi n/\omega$. For large $m$ the integral can then be evaluated semiclassically. For small circles and weak fields, $\omega L=L/2r_0 < 1$, the leading stationary point is real,
\begin{equation}
\beta_0=\frac{1}{\omega}\sin^{-1}\left(\omega L\right)
\end{equation}
and its contribution to the integral is free from ambiguities associated with the initial choice of contour.  Fig.~\ref{fig:beta0} shows the instanton with $\beta=\beta_0$ taken from the family plotted in Fig.~\ref{fig:sampletraj}. Expanding for $\omega L \ll 1$ gives
 \begin{equation}
Z^L_1/V^5=-\frac{m^2}{2\pi^2L^3}e^{-mL\left(1-\frac{1}{4}L^2\omega^2+\calO(\omega^4L^4)\right)}\left(1+\frac{1}{4}L^2\omega^2+\calO(\omega^4L^4)\right)\cos(A_5 Le)\;.
\label{eq:ZL1}
 \end{equation}
For $\omega= 0$, Eq.~(\ref{eq:ZL1}) becomes the leading term in the familiar Wilson line effective potential~\cite{hosotani}, $V(A_5)\sim e^{-mL}\cos(A_5Le)$ for $mL\gg1$. The background electric field reduces the exponential suppression of the amplitude. 

Furthermore, for $\omega L<1$, the instanton contribution is real: the leading semiclassical effect does not directly describe pair production or other decay modes for the flux. It describes a virtual  process before and after which the flux has the same value. Nonetheless, the amplitude is connected with a rather different type of flux-reducing process occurring in the long-distance theory, when the instanton amplitude is suitably interpreted as a Wilsonian effective action. 

\begin{figure}[t!]
\begin{center}
\includegraphics[width=0.45\linewidth]{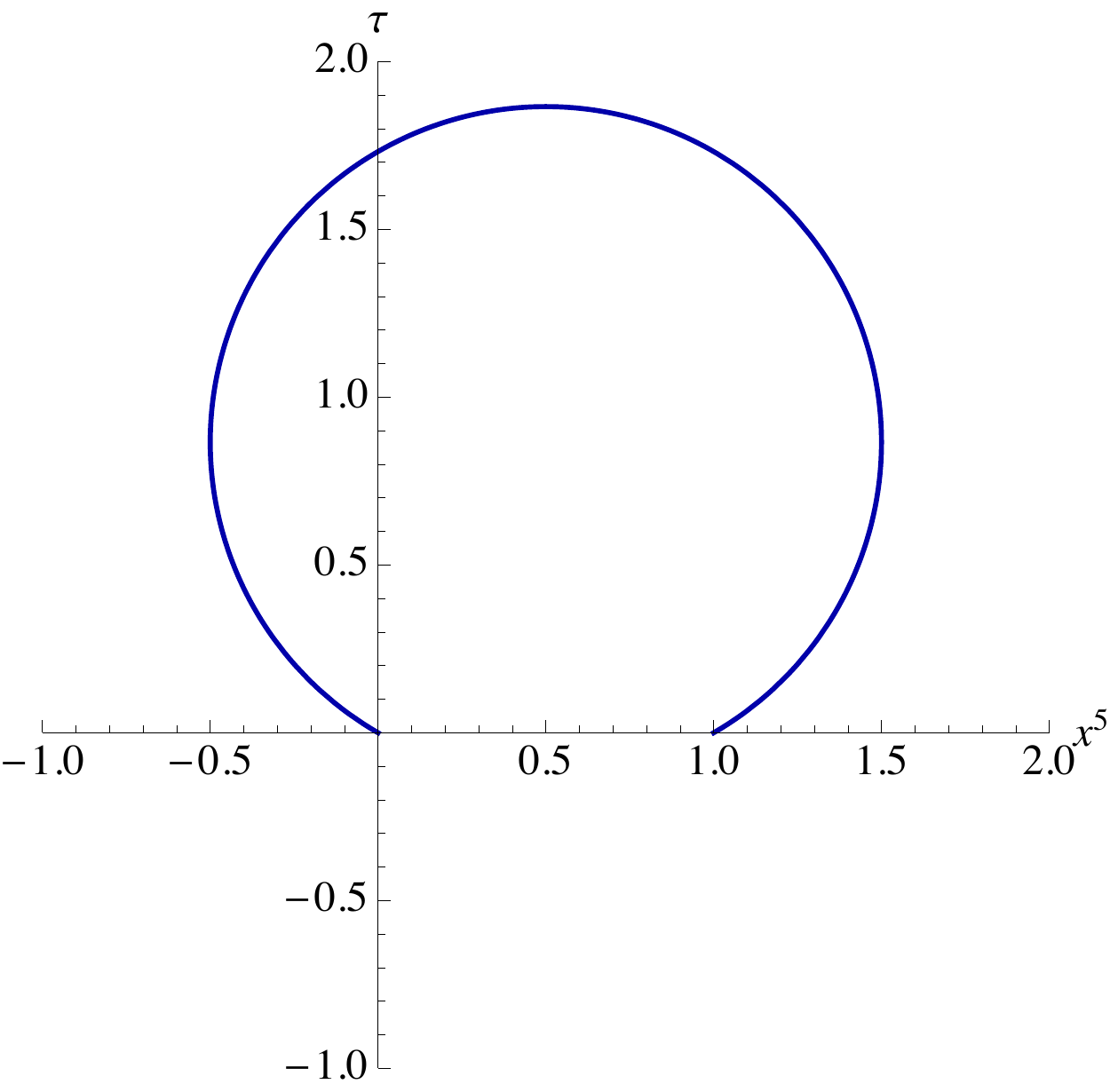}
\caption{Instanton with $\beta=\beta_1$, the saddle point of the second-lowest action in Eq.~(\ref{eq:ZL1}) and the first saddle with negative fluctuations. Parameters the same as in Fig.~\ref{fig:sampletraj}.}
\label{fig:beta1}
\end{center}
\end{figure}

In addition to $\beta_0$, Eq.~(\ref{eq:ZL1}) exhibits infinitely many saddle points of higher action.  The second saddle above the one shown in Fig.~\ref{fig:beta0} is given in Fig.~\ref{fig:beta1}. This instanton (and all subsequent even-numbered saddles) has a negative fluctuation eigenvalue corresponding to fluctuations in the $\beta$ direction, yielding an imaginary contribution to the effective action. The real part of the action is of order $\pi m r_0$ for small $\omega L$, associated with real pair production.

\section{$S_{eff}$ and Parametric Resonance}

It is convenient to work at distances larger than $L$, where $\calS_{eff}$ becomes a Wilsonian action for the four-dimensional gauge field and the periodic scalar zero mode of $A_5$, obtained by the substitutions
 \begin{align}
&A_5 Le\rightarrow \phi/f\;\;\;(f=1/eL)\nonumber\\
&E_5\rightarrow E_5\nonumber\\
&\phi\rightarrow E_5 t+\varphi(x,t),
\label{eq:rep}
 \end{align}
 where $E_5$ is a constant background velocity for $\phi$. $\phi$ is referred to as an axion. The meaning of the second line of Eq.~(\ref{eq:rep}) is that $E_5$ dependence already present in Eq.~(\ref{eq:ZL1}) through $\omega$ remains simply $E_5$, without introducing any $\dot\varphi$ terms. This truncation is necessary for the consistency of the derivative expansion. Furthermore, we will see shortly that we are interested in Fourier modes of $\varphi$ with momenta of order $eE_5L$. Consistently retaining the background $E_5=\dot A_5$ without computing $\partial_\mu \varphi$ terms in the effective action requires $e E_5L\varphi\lesssim E_5 \Rightarrow \varphi \lesssim f$. In other words, we must expand in small fluctuations $\varphi/f$.
 
Thus, at leading order in the derivative and semiclassical expansions and to quadratic order in $\varphi$, the scalar effective Lagrangian is
\begin{equation}
\calL_{eff,4d}(\varphi)=\frac{1}{2}(\partial\varphi)^2- \frac{m^2}{2\pi^2L^2}\left(\epsilon\frac{\varphi}{f}\sin\left(\Omega t\right)-\frac{1}{2}\left(\frac{\varphi}{f}\right)^2\epsilon\cos\left(\Omega t\right)\right),
\label{eq:Leff}
\end{equation}
where
\begin{equation}
\epsilon\equiv e^{-mL\left(1-\frac{1}{4}L^2\omega^2\right)},\;\;\;\;\;\Omega \equiv eE_5L,
 \end{equation}
retaining only the leading term in $\omega L$ in the exponent.

The linear term in $\varphi$ corrects the classical motion of the homogeneous mode. It is not important for our purposes and will subsequently be neglected. The quadratic term captures the leading effect responsible for the dissipation of flux in the effective theory. Non-exponentially-suppressed Euler-Heisenberg terms can be omitted because they are are flux-preserving, while real pair creation of the original charged fields has a much larger $\calO(m^2)$ classical action and cannot be described in the EFT. 

It is worth noting that it was not necessary to completely dimensionally reduce. Kaluza-Klein modes of $A_5$ up to level $n\sim mL$ could be retained in the effective theory below $m$, and modes up to $n\sim eE_5 L^2$ could play a role in the following discussion of dissipation. The reduction simplifies the analysis and is sufficient for  fields weak compared to the compactification scale, $eE_5L^2<1$, but in principle it could be extended to include additional modes.

In  backgrounds of constant $\dot \phi$, the mass term in~(\ref{eq:Leff}) is time-dependent, leading to a parametric resonance instability~\cite{traschenbrandenberger}. Amplitudes for modes of the axion with wavevectors $|k|\propto\dot \phi$ grow exponentially in time. The rate of the energy transfer to these modes may be obtained in several different ways. For nonzero $k$, the fluctuation equation of motion is 
\begin{equation}
\ddot\varphi(k,t)+k^2\varphi(k,t)+\epsilon\Lambda^2\cos(\Omega t)\varphi(k,t)=0
\end{equation}
where the scale $\Lambda^2\equiv \frac{m^2}{2\pi^2f^2L^2}=\frac{e^2m^2}{2\pi^2}$ is introduced to simplify notation. The $\epsilon$ term is a perturbation. Its first-order effect is to mix waves of frequency $\omega_k=\pm|k|$ with waves of frequency $\omega_k\pm\Omega$. For generic $|k|,$ the mixing is small and the solutions are stable. However, for $|k|\approx \Omega/2$, degenerate positive and negative frequency modes $\omega_k\approx \pm \Omega/2$ mix with each other, and the mixing has a more substantial effect. To analyze this case, we can restrict to deviations of order $\epsilon$ around $|k|=\Omega/2$,
\begin{align}
k^2\rightarrow (\Omega/2+&\epsilon \delta k)^2
\end{align}
and take the ansatz
\begin{align}
\varphi\rightarrow e^{i \left(\Omega/2+\epsilon\delta\omega_k\right)t} + c_0 e^{i \left(-\Omega/2+\epsilon\delta\omega_k\right)t}&+\epsilon c_+ e^{3i\Omega t/2 }+\epsilon c_- e^{-3i\Omega t/2 }.
\end{align}
This ansatz allows the waves with zeroth-order frequencies $\omega_k=\pm\Omega/2$ to mix with each other at zeroth order, and with modes of frequency $\pm3\Omega/2$ at first order. The equation of motion can then be solved to $\calO(\epsilon)$ for the coefficients $c_{0,+,-}$ and the frequency shift. The result for the frequency shift is
\begin{align}
\epsilon\delta\omega_k=\pm\frac{\epsilon}{2\Omega}\sqrt{-\Lambda^4+4\Omega^2 \delta k^2}\;.
\end{align}
In the band $|\epsilon\delta k|<\epsilon\Lambda^2/2\Omega$, $\delta\omega_k$ is imaginary, and some of the modes grow exponentially with $t$. For example, in the center of the band,
\begin{align}
\varphi\sim e^{i\Omega t /2 + \epsilon(\Lambda^2/2\Omega)t}.
\end{align}
Consequently, the energy in the first band grows approximately as
\begin{equation}
\int d^3k\,\omega_k \langle|\varphi(k,t)|^2\rangle \sim (\Omega)^2 \left(\epsilon \frac{\Lambda^2}{\Omega}\right) \langle|\varphi(k,0)|^2\rangle e^{\epsilon(\Lambda^2/\Omega)t}\;.
\end{equation}
At higher orders in $\epsilon$, instabilities appear in frequency bands centered on every $n\Omega/2$.

These results have a simple general description. In the absence of charged matter, electric flux is conserved. Electric fields along a compact direction are constant as a result of improper gauge transformations, a global shift symmetry acting as $A_5\rightarrow A_5+c/L,\;\;c\sim c+2\pi$. Charged matter breaks the global symmetry, so some flux non-conserving processes are expected to survive in a long-distance effective theory. The leading process is parametric resonance, driven by the ordinary Wilson line effective potential terms $\sim e^{-mL}\cos(nA_5)$. The worldline instanton  captures flux-dependent corrections that reduce the exponent. This reduction was also noted in~\cite{brown}, although the instanton and antiinstanton in Fig.~\ref{fig:beta0} were treated as a single event, and so the result did not contain the harmonic terms in $A_5$ that are critical to flux dissipation. It was also suggested in~\cite{brown} that the instanton describes annihilation of virtual electrons into photons, a process which is not captured by the leading semiclassical approximation. 
Instead, dissipation arises in the EFT by production of scalar quanta.

The discussion above is relevant in $d>1$ spatial dimensions. The 1+1 dimensional case is somewhat different, but also of interest. Here the electric field acts as a $\theta$-term~\cite{coleman}, and Schwinger pair production describes tunneling between different metastable branches, analogous to phenomena arising in various 4d gauge theories~\cite{wittenlargen}.  
Dissipation by parametric resonance, however, does not occur when space is compactified on a small circle. In this case the effective theory at distances long compared to $L$ is  the quantum mechanics of a 1D rotor $\phi(t)$. Flux/angular momentum $\sim \dot\phi$ is quantized and conserved at $\calO(\epsilon^0)$. The contribution to the Hamiltonian at $\calO(\epsilon)$ does not commute with the angular momentum and describes mixing of different flux states.

\section{Finite Temperature}
The instantons studied above in the case of a spatial dimension have also been suggested to be relevant at finite temperature~\cite{MO}.  Here I will comment briefly on the relation of the finite temperature and spatial circle instantons, without attempting a complete discussion of the thermal problem.

Locally, arcs of the critical circle still solve the Euclidean equations of motion, regardless of the boundary conditions. However, with thermal boundary conditions $\tau\sim\tau+L$ and electric flux pointing in a noncompact spatial direction $z$, trajectories of nonzero winding number  are no longer stationary points of the Euclidean action. This includes worldlines analogous to Figs.~\ref{fig:beta0} and~\ref{fig:beta1}, swapping $x_5\rightarrow \tau,~\tau\rightarrow z$. The $\beta$ action is a linear function of a collective coordinate $z_0$  describing the starting and ending point of the worldline in the noncompact flux direction. For example, the action of a trajectory analogous to Fig.~\ref{fig:beta0} would be
\begin{align}
\calA_0=e E_z L z_0+\frac{1}{4}e E_z L^2\cot\left(\omega\beta\right)\;.
\end{align}

\begin{figure}[t!]
\begin{center}
\includegraphics[width=0.5\linewidth]{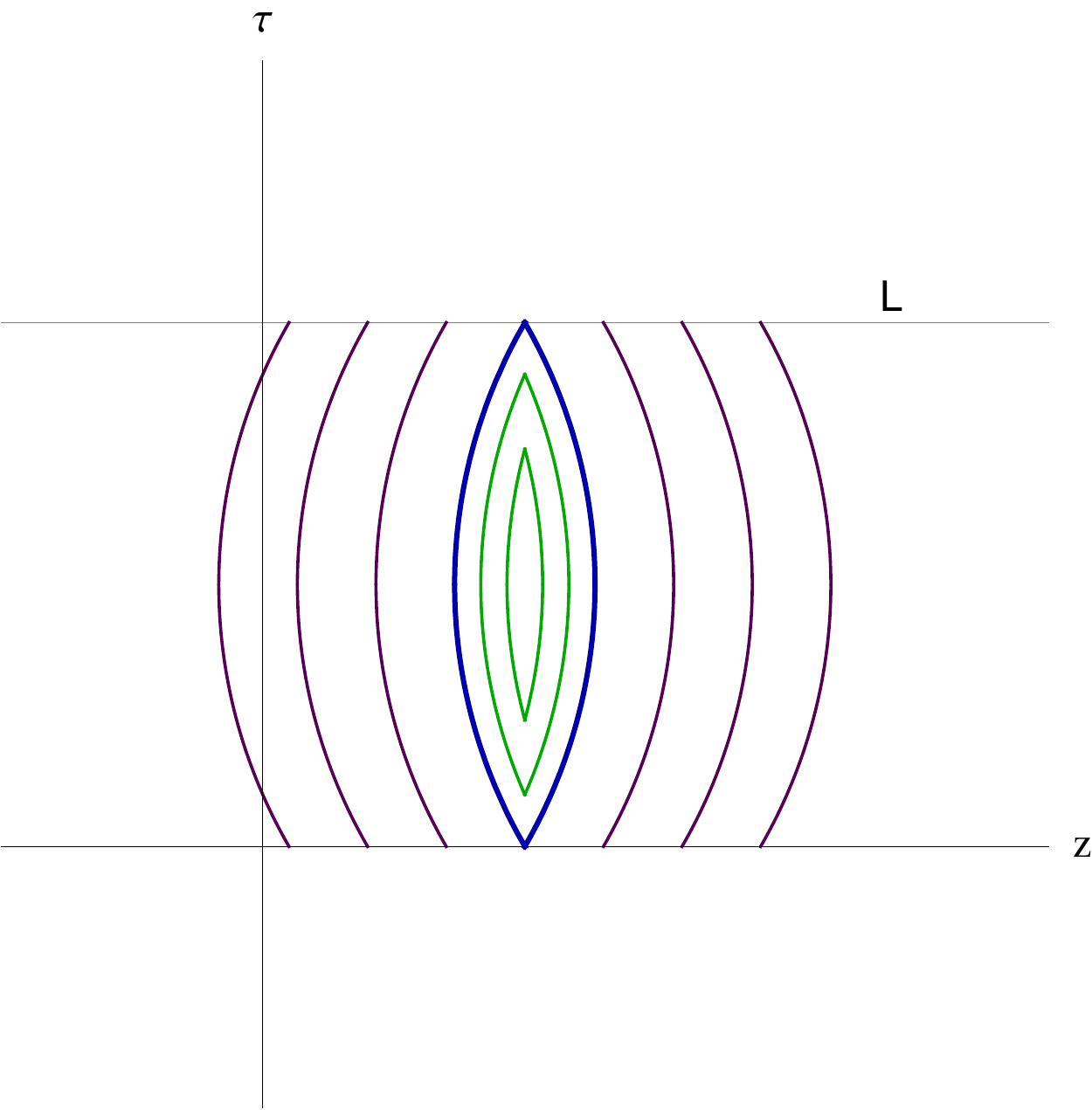}
\caption{A family of trajectories exhibiting the distinguished nature of the critical lemon worldline (blue). Smaller lemons (green) locally solve the equations of motion, but they are not true stationary points because the action rises monotonically in the direction of the critical lemon. Once the lemon grows to the critical size, its two sides can split apart, opening a new direction in configuration space. In this direction the action changes linearly with separation. These features suggest the presence of a saddle point with a negative mode near the critical lemon.}
\label{fig:negativemode}
\end{center}
\end{figure} 

Consequently, in contrast to the spatial case, the thermal semiclassical approximation must be based on trajectories of zero winding. Furthermore, if they are to be different from the ordinary zero-temperature solutions, the trajectories must use the thermal circle in an essential way. One possibility is that interaction with the image charges produces new classes of solutions. Such effects are of order $e^2$, but can be non-negligible in $d>2$ dimensions when charges approach each other. In particular, the weak-field expansion  breaks down in regions where the worldlines intersect or exhibit kink discontinuities. It is less clear that $e^2$ effects can be important in $1+1$ dimensions for $E\gg e$. Another possibility is that discontinuous trajectories of net winding number zero, like instanton-antiinstanton events in the case of the spatial circle, play a role.

In~\cite{brown}, it was argued that a worldline corresponding to the fusing of the instanton and antiinstanton in Fig.~\ref{fig:beta0} describes thermally-assisted pair creation. This trajectory, referred to as the ``lemon instanton," was studied further in four dimensions in~\cite{GR1}, including the role of short-distance interactions in modifying the worldline.  In~\cite{brown}, the lemon was also shown to reproduce the exponential suppression obtained by minimizing the sum of  Boltzmann and WKB tunneling exponents, treating the tunneling process as a relativistic barrier penetration problem with linear potential.

The lemon instanton can also be described as (the boundary of) the overlap region between two circles of Schwinger radius $r_0$, separated by a distance $d=d_L$ so that the lemon ``just fits" inside the thermal circle, $d_L/2\equiv\sqrt{r_0^2-L^2/4}$. In this light, the trajectory appears puzzling: small variations of $d$  change the action at leading order in the variation. The only stationary point of the $\calO(e)$ action appears to be $d=0$, where the circles degenerate. At higher order in $e$, different parts of the worldlines can interact; furthermore, with a compact direction, trajectories larger than $L$ begin to overlap, increasing the effects of interactions. However, it is unclear whether these effects are important in $1+1$ dimensions, where the theory is otherwise simplest.

Another feature of the lemon is that it marks a singular point in configuration space where continuous trajectories, for example lemons of various $d$, can first split into discontinuous  pairs of opposite winding number. A family of configurations exhibiting this property is sketched in Fig.~\ref{fig:negativemode}; among them, the critical lemon with $d=d_L$ maximizes the action. Smaller lemons (larger $d$) cannot split; this direction in field space opens up sharply when $d\rightarrow d_L$, allowing the action to lower again. This picture suggests that in $1+1$ dimensions the lemon is indeed a distinguished trajectory, and may be near to a genuine saddle point with a negative fluctuation eigenvalue once singularities are smoothed out by ultraviolet effects.

\section{Bounds on Field Excursions}

Theories of light axions may be subject to theoretical constraints including the weak gravity and swampland conjectures~\cite{swampland1,swampland2,banksdinefoxgorbatov,wgc,Obied:2018sgi,Agrawal:2018own}, and perhaps related to these constraints, there can be limitations on physically allowed objects and dynamical processes, particularly involving large excursions of the scalar in spacetime~\cite{Banks:2004im}. For example, large stationary excursions can collapse into black holes~\cite{nicolis} or destabilize Kaluza-Klein spacetimes~\cite{draperfarkas}, while large axion excursions around cosmic strings cause them to inflate~\cite{vilenkin,draperetal,Hebecker:2017wsu}. In this light we can ask whether parametric resonance places limitations on axion excursions, when the axion arises from a higher-dimensional U(1).

An axion with action~(\ref{eq:Leff}) and decay constant $f$, moving with initial velocity $\dot\phi$, fragments into inhomogeneous modes in a timescale of order 
\begin{align}
\log(t)\sim\log(1/\epsilon)\sim mL= m/ef.
\label{eq:logt}
\end{align}
In comparison, a scalar with zero potential but subject to Hubble friction decelerates as
$\ddot\phi\lesssim-\dot\phi^2/M_p$, saturated when the scalar is the dominant source of energy. In a time $t$ it moves a proper distance $\Delta\phi\lesssim M_p\log(t)$ in field space. Inserting~(\ref{eq:logt}), we can estimate that the slowly fragmenting scalar can move a distance of order
\begin{align}
\Delta\phi\lesssim \frac{M_p m}{ef}
\end{align}
before dissipating. Imposing a weak-gravity relation $m/e<M_p$, we obtain
\begin{align}
\Delta\phi\lesssim \frac{M_p^2}{f}.
\label{eq:deltaphi}
\end{align}
Weak gravity also requires $f<M_p$, so no two distinct points in the axion field space are separated by a transplanckian distance. But we see that the bound~(\ref{eq:deltaphi}) carries additional information: even with subplanckian field range, the distance physically traversable by a homogeneous field is bounded.

\vskip 1cm
\noindent
{\bf Acknowledgements:} I thank L. Sorbo, A. Nahum, and G. Dunne for discussions. This work was supported by NSF grant PHY-1719642.

\bibliography{schwinger}{}

\providecommand{\href}[2]{#2}\begingroup\raggedright\begin{thebibliography}{10}

\bibitem{Schwinger:1951nm}
J.~S. Schwinger, ``{On gauge invariance and vacuum polarization},''
  \href{http://dx.doi.org/10.1103/PhysRev.82.664}{{\em Phys. Rev.} {\bfseries
  82} (1951) 664--679}.
[,116(1951)].

\bibitem{brown}
A.~R. Brown, ``{Schwinger pair production at nonzero temperatures or in compact
  directions},''
\href{http://arxiv.org/abs/1512.05716}{{\ttfamily arXiv:1512.05716 [hep-th]}}.

\bibitem{MO}
L.~Medina and M.~C. Ogilvie, ``{Schwinger Pair Production at Finite
  Temperature},'' \href{http://dx.doi.org/10.1103/PhysRevD.95.056006}{{\em
  Phys. Rev.} {\bfseries D95} no.~5, (2017) 056006},
\href{http://arxiv.org/abs/1511.09459}{{\ttfamily arXiv:1511.09459 [hep-th]}}.

\bibitem{GR1}
O.~Gould and A.~Rajantie, ``{Thermal Schwinger pair production at arbitrary
  coupling},'' \href{http://dx.doi.org/10.1103/PhysRevD.96.076002}{{\em Phys.
  Rev.} {\bfseries D96} no.~7, (2017) 076002},
\href{http://arxiv.org/abs/1704.04801}{{\ttfamily arXiv:1704.04801 [hep-th]}}.

\bibitem{GR2}
O.~Gould, A.~Rajantie, and C.~Xie, ``{Worldline sphaleron for thermal Schwinger
  pair production},''
\href{http://arxiv.org/abs/1806.02665}{{\ttfamily arXiv:1806.02665 [hep-th]}}.

\bibitem{arun}
M.~Korwar and A.~M. Thalapillil, ``{Finite temperature Schwinger pair
  production in coexistent electric and magnetic fields},''
\href{http://arxiv.org/abs/1808.01295}{{\ttfamily arXiv:1808.01295 [hep-th]}}.

\bibitem{kleinert}
H.~Kleinert,
``{Path Integrals in Quantum Mechanics, Statistics, Polymer Physics, and
  Financial Markets},''.

\bibitem{hosotani}
Y.~Hosotani, ``{Dynamical Mass Generation by Compact Extra Dimensions},''
\href{http://dx.doi.org/10.1016/0370-2693(83)90170-3}{{\em Phys. Lett.}
  {\bfseries 126B} (1983) 309--313}.

\bibitem{traschenbrandenberger}
J.~H. Traschen and R.~H. Brandenberger, ``{Particle Production During
  Out-of-equilibrium Phase Transitions},''
\href{http://dx.doi.org/10.1103/PhysRevD.42.2491}{{\em Phys. Rev.} {\bfseries
  D42} (1990) 2491--2504}.

\bibitem{coleman}
S.~R. Coleman, ``{More About the Massive Schwinger Model},''
\href{http://dx.doi.org/10.1016/0003-4916(76)90280-3}{{\em Annals Phys.}
  {\bfseries 101} (1976) 239}.

\bibitem{wittenlargen}
E.~Witten, ``{Large N Chiral Dynamics},''
\href{http://dx.doi.org/10.1016/0003-4916(80)90325-5}{{\em Annals Phys.}
  {\bfseries 128} (1980) 363}.

\bibitem{swampland1}
C.~Vafa, ``{The String landscape and the swampland},''
\href{http://arxiv.org/abs/hep-th/0509212}{{\ttfamily arXiv:hep-th/0509212
  [hep-th]}}.

\bibitem{swampland2}
H.~Ooguri and C.~Vafa, ``{On the Geometry of the String Landscape and the
  Swampland},'' \href{http://dx.doi.org/10.1016/j.nuclphysb.2006.10.033}{{\em
  Nucl. Phys.} {\bfseries B766} (2007) 21--33},
\href{http://arxiv.org/abs/hep-th/0605264}{{\ttfamily arXiv:hep-th/0605264
  [hep-th]}}.

\bibitem{banksdinefoxgorbatov}
T.~Banks, M.~Dine, P.~J. Fox, and E.~Gorbatov, ``{On the possibility of large
  axion decay constants},''
  \href{http://dx.doi.org/10.1088/1475-7516/2003/06/001}{{\em JCAP} {\bfseries
  0306} (2003) 001},
\href{http://arxiv.org/abs/hep-th/0303252}{{\ttfamily arXiv:hep-th/0303252
  [hep-th]}}.

\bibitem{wgc}
N.~Arkani-Hamed, L.~Motl, A.~Nicolis, and C.~Vafa, ``{The String landscape,
  black holes and gravity as the weakest force},''
  \href{http://dx.doi.org/10.1088/1126-6708/2007/06/060}{{\em JHEP} {\bfseries
  06} (2007) 060},
\href{http://arxiv.org/abs/hep-th/0601001}{{\ttfamily arXiv:hep-th/0601001
  [hep-th]}}.

\bibitem{Obied:2018sgi}
G.~Obied, H.~Ooguri, L.~Spodyneiko, and C.~Vafa, ``{De Sitter Space and the
  Swampland},''
\href{http://arxiv.org/abs/1806.08362}{{\ttfamily arXiv:1806.08362 [hep-th]}}.

\bibitem{Agrawal:2018own}
P.~Agrawal, G.~Obied, P.~J. Steinhardt, and C.~Vafa, ``{On the Cosmological
  Implications of the String Swampland},''
  \href{http://dx.doi.org/10.1016/j.physletb.2018.07.040}{{\em Phys. Lett.}
  {\bfseries B784} (2018) 271--276},
\href{http://arxiv.org/abs/1806.09718}{{\ttfamily arXiv:1806.09718 [hep-th]}}.

\bibitem{Banks:2004im}
T.~Banks, ``{Strings in a Landscape},'' in {\em {String theory: From gauge
  interactions to cosmology}}, pp.~3--7.
\newblock 2004.
\newblock
\url{http://link.springer.com/chapter/10.1007/1-4020-3733-3_1}.
\newblock

\bibitem{nicolis}
A.~Nicolis, ``{On Super-Planckian Fields at Sub-Planckian Energies},''
  \href{http://dx.doi.org/10.1088/1126-6708/2008/07/023}{{\em JHEP} {\bfseries
  07} (2008) 023},
\href{http://arxiv.org/abs/0802.3923}{{\ttfamily arXiv:0802.3923 [hep-th]}}.

\bibitem{draperfarkas}
P.~Draper and S.~Farkas. {To appear}.

\bibitem{vilenkin}
A.~Vilenkin, ``{Topological inflation},''
  \href{http://dx.doi.org/10.1103/PhysRevLett.72.3137}{{\em Phys. Rev. Lett.}
  {\bfseries 72} (1994) 3137--3140},
\href{http://arxiv.org/abs/hep-th/9402085}{{\ttfamily arXiv:hep-th/9402085
  [hep-th]}}.

\bibitem{draperetal}
M.~J. Dolan, P.~Draper, J.~Kozaczuk, and H.~Patel, ``{Transplanckian Censorship
  and Global Cosmic Strings},''
  \href{http://dx.doi.org/10.1007/JHEP04(2017)133}{{\em JHEP} {\bfseries 04}
  (2017) 133},
\href{http://arxiv.org/abs/1701.05572}{{\ttfamily arXiv:1701.05572 [hep-th]}}.

\bibitem{Hebecker:2017wsu}
A.~Hebecker, P.~Henkenjohann, and L.~T. Witkowski, ``{What is the Magnetic Weak
  Gravity Conjecture for Axions?},''
  \href{http://dx.doi.org/10.1002/prop.201700011}{{\em Fortsch. Phys.}
  {\bfseries 65} no.~3-4, (2017) 1700011},
\href{http://arxiv.org/abs/1701.06553}{{\ttfamily arXiv:1701.06553 [hep-th]}}.

\end{thebibliography}\endgroup
\bibliographystyle{utphys}

\end{document}